\documentclass[%
 reprint,
 amsmath,amssymb,
 aps,
floatfix,
]{revtex4-1}

\usepackage{graphicx}
\usepackage{dcolumn}
\usepackage{bm}
\usepackage{amsmath}
\usepackage{amssymb}
\usepackage{xcolor}
\usepackage{hyperref}
\usepackage[capitalise]{cleveref}

\crefname{chapter}{chap.}{chap.}
\crefname{section}{Sec.}{Sec.}
\Crefname{chapter}{Chapter}{Chapters}
\Crefname{section}{Section}{Sections}

\begin{document}

\preprint{APS/123-QED}

\title{Early dark energy constraints on growing neutrino quintessence cosmologies}

\author{Finlay Noble Chamings}
 \email{finlay.noblechamings@nottingham.ac.uk}
\author{Anastasios Avgoustidis}%
\author{Edmund J. Copeland}
\author{Anne M. Green}
\affiliation{%
 School of Physics and Astronomy, University of Nottingham
 \\Nottingham NG7 2RD, United Kingdom
}%

\author{Baojiu Li}
\affiliation{
 Institute for Computational Cosmology, Department of Physics, Durham University
 \\Durham DH1 3LE, United Kingdom
}%

\date{\today}

\begin{abstract}
We investigate cosmological models in which dynamical dark energy consists of a scalar field whose present-day value is controlled by a coupling to the neutrino sector. The behaviour of the scalar field depends on three functions: a kinetic function, the scalar field potential, and the scalar field-neutrino coupling function. We present an analytic treatment of the background evolution during radiation- and matter-domination for exponential and inverse power law potentials, and find a relaxation of constraints compared to previous work on the amount of early dark energy in the exponential case. We then carry out a numerical analysis of the background cosmology for both types of potential and various illustrative choices of the kinetic and coupling functions. By applying bounds from Planck on the amount of early dark energy, we are able to constrain the magnitude of the kinetic function at early times.

\end{abstract}

\pacs{Valid PACS appear here}

\maketitle

\section{\label{sec:intro}Introduction}

Since the acceleration of the expansion of the Universe was discovered by the study of Type Ia supernovae~\cite{Riess:1998,Schmidt:1998ys}, uncovering the cause of this acceleration has become one of the most important problems in cosmology. The so-called concordance model of cosmology, or $\Lambda$CDM, is the simplest cosmological model that gives a good fit to the Type Ia supernovae data~\cite{Riess:2016jrr} as well as the cosmic microwave background~\cite{Planck2018VI} and large scale structure~\cite{SDSS14}. In $\Lambda$CDM the late-time universal expansion is caused by a cosmological constant denoted by $\Lambda$. 

Despite its experimental successes, however, $\Lambda$CDM suffers from a number of theoretical problems, notably the coincidence problem~\cite{whynow}, the fine-tuning problem~\cite{cosmoconstprob} and the difficulty of explaining what fundamentally gives rise to the cosmological constant~\cite{Polchinski:2006gy}. This has motivated the study of a number of possible explanations of the observed late-time expansion not involving a cosmological constant, including modified gravity models~\cite{modgravreview} and dynamical dark energy~\cite{Copeland:2006wr}. Quintessence is a simple form of dynamical dark energy which consists of a single scalar field $\varphi$ whose evolution is slow enough to give rise to a negative pressure, hence accelerated expansion and late time domination of the energy density~\cite{Copeland:2006wr,Tsujikawa:2013fta}.

One novel approach to solving the coincidence problem is to give $\varphi$ a non-standard coupling to neutrinos such that the neutrino mass is $\varphi$-dependent. The scalar field evolves according to a scaling solution~\cite{Wetterichdilatation,Copelandscaling} that ends as the neutrinos become non-relativistic and, through the scalar-neutrino coupling, provide a force on $\varphi$ that effectively stops it from evolving. At this point $\varphi$ is potential-dominated and acquires a negative equation of state, giving rise to a period of dark energy domination. Such models are known as ``growing neutrino quintessence" models~\cite{GNQ1,GNQ2}, so called because the neutrino mass grows as the scalar field evolves. Growing neutrino quintessence models are closely related to ``mass varying neutrino'' (MaVaN) models, which also involve an interaction between dark energy and the neutrino mass~\cite{Fardon_2004, Afshordi:2005ym, PhysRevLett.93.091801, PhysRevD.72.123523, PhysRevD.73.083515, Spitzer:2006hm, Takahashi_2006, 2013A&A...560A..53L}. As well as providing the right conditions for the scalar field to play the role of dark energy, the neutrino-scalar coupling can give rise to an attractive fifth-force acting on the neutrinos that is much stronger than gravity. This force gives rise to non-linear ``neutrino lumps'' on large scales, which have been extensively studied using linear approximation~\cite{Mota:2008nj,PhysRevD.82.123001}, N-body simulations~\cite{2011MNRAS.418..214B,PhysRevD.85.123010,PhysRevD.87.043519,Ayaita:2014una,Fuhrer:2015xya,Casas:2016duf}, spherical collapse~\cite{PhysRevD.82.103516}, and other methods~\cite{PhysRevD.81.063525,Nunes:2011mw,Brouzakis_2011}. The effect these neutrino lumps have on the cosmological history depends on the masses of neutrinos. As found in Ref.~\cite{Casas:2016duf}, for large neutrino masses the neutrino lumps can be stable and can lead to significant backreaction effects. For smaller neutrino masses, however, the neutrino lumps are unstable; they form and dissociate periodically such that backreaction effects are small.

In this work we consider a number of growing neutrino quintessence models, investigating the effects of changing the scalar field kinetic and potential terms and the scalar-neutrino coupling on the background evolution of the scalar field. We study radiation and matter domination analytically, and numerically solve the full system of background equations, in order to gain insight into the robustness of growing neutrino quintessence to variations in the model parameters.

This paper is organised as follows: in \cref{sec:model} we describe a particular growing neutrino quintessence model proposed by Wetterich in Ref.~\cite{wetterichinfquint} and state the equations of motion; in \cref{sec:anal} we present approximate analytic solutions to the model in Ref.~\cite{wetterichinfquint} and a related model;
in \cref{sec:results} we describe our numerical analysis, present the numerical results and discuss the constraints on model parameters; and finally we summarise in \cref{sec:conc}.

\section{\label{sec:model}The Model}

In Ref.~\cite{wetterichinfquint}, Wetterich proposed a unified model of inflation and quintessence. The model consists of a crossover from a past, ultraviolet fixed point to a future, infrared fixed point. Each of these regions is approximately scale invariant and corresponds to inflation and dark energy domination respectively. This is achieved by introducing a scalar field, which in a particular choice of conformal frame acts as a variable Planck mass. The scalar field is coupled to the neutrinos in such a way as to produce growing neutrino quintessence~\cite{GNQ1,GNQ2}, with the transition to dark energy domination being driven by the ``trigger" event of the neutrinos becoming non-relativistic.

In this work we consider the crossover region, since 
it can be constrained by presently available cosmological data.
According to the model in Ref.~\cite{wetterichinfquint}, radiation domination, matter domination and the present transition to dark energy domination are all part of the crossover.
In this period the model effectively reduces to a scalar field minimally coupled to both gravity and the matter sector, with the only exception being the coupling to the neutrino sector. For this reason it is convenient to work in the Einstein frame, in which the model is described by the following action:
\begin{multline}
\label{eq:action}
  S = \int \text{d}^4x \sqrt{-g} \left[ \frac{1}{2} M_\text{P}^2 R - \frac{1}{2} k^2(\varphi) \nabla_\mu\varphi \nabla^\mu\varphi - V(\varphi) \right]
  \\+ S_\text{m}[\Psi_\text{m}, g_{\mu\nu}] + S_\gamma[\Psi_\gamma, g_{\mu\nu}] 
  \\+ S_\nu[\Psi_\nu, C(\varphi)^2 g_{\mu\nu}]\,,
\end{multline}
where $M_\text{P}$ is the reduced Planck mass, $k^2(\varphi)$, $V(\varphi)$ and $C(\varphi)$ are respectively the kinetial, potential and neutrino-scalar coupling function and must be specified in order to choose a particular model. $\Psi_\text{m}$, $\Psi_\gamma$ and $\Psi_\nu$ correspond to the matter, radiation and neutrino fields respectively. The key feature of growing neutrino quintessence models is the function $C(\varphi)$ which couples neutrinos to the scalar field and effectively gives the neutrinos a time-dependent mass given by:
\begin{equation}
m_\nu(\varphi) = \bar{m}_\nu C(\varphi)\,,
\end{equation}
where $\bar{m}_\nu$ is a mass scale. For simplicity we take all neutrino masses to be equal. It is often convenient to work in terms of the dimensionless function: 
\begin{equation}
\label{eq:betadef}
\beta(\varphi) \equiv - M_\text{P}\frac{\text{d}\ln C(\varphi)}{\text{d}\varphi}\,.
\end{equation}

By varying the action, \cref{eq:action}, with respect to the metric $g_{\mu\nu}$ one obtains the gravitational field equations:
\begin{multline}\label{Einstein-tensor}
G_{\mu\nu} = \frac{1}{M_\text{P}^2}T_{\mu\nu} + \frac{1}{M_\text{P}^2}\biggl[ k^2(\varphi)\nabla_\mu\varphi\nabla_\nu\varphi 
\\- \frac{1}{2}k^2(\varphi)\nabla_\rho\varphi\nabla^\rho\varphi g_{\mu\nu} - V(\varphi) g_{\mu\nu}\biggr]\,,
\end{multline}
and varying with respect to the scalar field $\varphi$ yields the scalar field equation of motion:
\begin{equation}\label{KGeqn}
-k^2\nabla_\mu\nabla^\mu\varphi - \frac{1}{2}\frac{\text{d}k^2}{\text{d}\varphi}\nabla_\mu\varphi\nabla^\mu\varphi + \frac{\text{d}V}{\text{d}\varphi} + \beta \frac{T^{(\nu)\mu}_\mu}{M_\text{P}} = 0\,.
\end{equation}
Here $T_{\mu\nu}$ is the total stress-energy-momentum tensor of all species apart from the scalar field and $T_{\mu\nu}^{(\nu)}$ is the stress-energy-momentum tensor for the neutrinos.

If we assume a spatially-flat Friedmann-Lema\^{i}tre-Robertson-Walker metric of the form
\begin{equation}
\text{d}s^2 = -\text{d}t^2 + a(t)^2\delta_{ij}\text{d}x^i\text{d}x^j\,,
\end{equation}
and assume the scalar field is homogeneous, then \cref{Einstein-tensor,KGeqn} become:
\begin{equation}
H^2 = \frac{\rho}{3M_\text{P}^2}\,,
\end{equation}
\begin{equation}
\dot{H} = -\frac{\rho+p}{2M_\text{P}^2}\,,
\end{equation}
and
\begin{equation}
\label{eq:SFE}
\ddot{\varphi} + 3H\dot{\varphi} + \frac{1}{2k^2}\frac{\text{d}k^2}{\text{d}\varphi}\dot{\varphi}^2 + \frac{1}{k^2}\frac{\text{d}V}{\text{d}\varphi} - \frac{\beta}{M_\text{P}}(\rho_\nu - 3p_\nu) = 0\,,
\end{equation}
where $H\equiv\dot{a}/a$ is the Hubble parameter, dots denote differentiation with respect to time, $\rho = \rho_\text{m} + \rho_\nu + \rho_\gamma + \rho_\varphi$ and $p = p_\text{m} + p_\nu + p_\gamma + p_\varphi$ are the energy density and pressure of all species. The energy density and pressure of the homogeneous scalar field are defined as:
\begin{equation}
\label{eq:rhophidef}
\rho_\varphi = \frac{k^2}{2}\dot{\varphi}^2 + V\,,
\end{equation}
\begin{equation}
\label{eq:pphidef}
p_\varphi = \frac{k^2}{2}\dot{\varphi}^2 - V\,.
\end{equation}

Matter and radiation obey the usual conservation equations: $\dot{\rho}_\text{m} + 3H\rho_\text{m} = 0$ and $\dot{\rho}_\gamma + 4H\rho_\gamma = 0$. However, the neutrinos obey a modified conservation equation due to their interaction with the scalar field:
\begin{equation}
\label{eq:rhonucons}
\dot{\rho}_\nu + 3H(\rho_\nu + p_\nu) = - \frac{\beta}{M_\text{P}}(\rho_\nu - 3p_\nu)\dot{\varphi}\,,
\end{equation}
\begin{equation}
\label{eq:rhophicons}
\dot{\rho}_\varphi + 3H(\rho_\varphi + p_\varphi) = \frac{\beta}{M_\text{P}}(\rho_\nu - 3p_\nu)\dot{\varphi}\,.
\end{equation}

For most of the Universe's history, the neutrinos are highly relativistic and $\rho_\nu - 3p_\nu \approx 0$ such that the scalar field and the neutrinos are effectively uncoupled and the scalar field energy density tracks that of the dominant species.
After the neutrinos become non-relativistic the coupling becomes important and, for large enough $|\beta|$, effectively stops the evolution of the scalar field by providing a force to counter that caused by the gradient of the potential in \cref{eq:SFE}.
As a result, the scalar field's energy density and pressure are dominated by the potential and the equation of state $w_\varphi \equiv p_\varphi/\rho_\varphi$ approaches $-1$, which is consistent with observations~\cite{Planck2018VI}.

\section{Approximate analytic solutions}
\label{sec:anal}

Under certain simplifying assumptions, it is possible to solve the scalar field equation, \cref{eq:SFE}, analytically. In this section we consider the behaviour of $\varphi$ before the neutrinos become non-relativistic, both for an exponential and an inverse power law potential. For the exponential case the scalar field evolves linearly with $N=\log(a)$ and there is an approximately constant fraction of early dark energy present. In the inverse power law case we find instead that $\log(\varphi)$ evolves linearly with $N$. For an analytic treatment of a related but distinct model, see Ref.~\cite{Hossain:2014xha}.

\subsection{Exponential potential}

First we consider a constant kinetic function $k^2(\varphi)=k^2_\text{c}=\text{const}$ and an exponential potential $V(\varphi) = M_\text{P}^4 \exp(-\alpha\varphi/M_\text{P})$, where $\alpha$ is a dimensionless parameter that determines the slope of the potential.
Before the neutrinos have become non-relativistic, the model exhibits a scaling solution whereby the energy density of the scalar field tracks that of the dominant species (radiation or matter, depending on the epoch) with the result that the energy density fraction of the scalar field is constant. 
It is convenient to introduce the energy density of the dominant species as $\rho_\text{d}$, which is equal to $\rho_\gamma + \rho_\nu$ in the radiation-dominated epoch and $\rho_\text{m}$ in the matter-dominated epoch. Here we are considering only the epoch in which neutrinos are highly relativistic, so the coupling between the neutrinos and the scalar field is effectively zero and neutrinos can be treated simply as radiation along with the photons. 

Sufficiently far from matter-radiation equality one can neglect whichever of matter and radiation is subdominant and write:
\begin{equation}
    \rho_\text{tot} = \rho_\text{d} + \rho_\varphi\,,
\end{equation}
where the energy density of the dominant species evolves as
\begin{equation}
\label{eq:rhod}
    \rho_\text{d} \propto \exp(-nN)\,,
\end{equation}
where $N=0$ at the present time and $n=4\,(3)$ for radiation (matter) domination. 
In the scaling solution,
\begin{equation}
\label{eq:rhop}
    \rho_\varphi \propto \exp(-nN)\,,
\end{equation}
and the (constant) energy density fraction of the scalar field is given by
\begin{equation}
\label{eq:Oscalsol}
\Omega_\varphi = \frac{nk^2_\text{c}}{\alpha^2}\,.
\end{equation}
The scalar field itself obeys the following particular solution of \cref{eq:SFE}
\begin{equation}
\label{eq:phiscalsol}
\varphi = M_\text{P}\frac{nN}{\alpha}+\hat{\varphi}\,,
\end{equation}
where $\hat{\varphi}$ is the value $\varphi$ would take at $N=0$ (though note that this bears no relation to realistic present-day values of $\varphi$ since at some point before $N=0$ the neutrinos become important and the scaling solution becomes invalid).

For a slowly-varying kinetic function $k^2(\varphi)$ we can expect behaviour that approximates this scaling solution. The procedure for quantifying the deviation from the scaling solution was demonstrated in Ref.~\cite{wetterichinfquint}, which in turn is based on a calculation in Ref.~\cite{Wetterich:1994bg}. When we carry out this procedure we find disagreement with the results of Ref.~\cite{wetterichinfquint}. The details of our calculation are laid out in \cref{appendix}. Here we simply present the results.

We find that the energy density of the scalar field obeys
\begin{equation}
\label{eq:EDE1}
    \Omega_\varphi = \frac{n k^2}{\alpha^2}(1-\bar{u})\,,
\end{equation}
which deviates from the exact scaling solution by a small quantity
\begin{equation}
\label{eq:constu}
    \bar{u} = \frac{M_\text{P}}{\alpha(1-\Omega_\varphi)}\frac{\text{d}\log k^2}{\text{d}\varphi}\,.
\end{equation}
This differs from the corresponding result in Ref.~\cite{wetterichinfquint} by a factor of $\Omega_\varphi$. As an example, we consider the particular kinetic function used in Ref.~\cite{wetterichinfquint}:
\begin{equation}
\label{eq:k}
    k_1^2(\varphi) = \frac{M_\text{P}\alpha}{2\kappa(\varphi-\bar{\varphi})}\,,
\end{equation}
where $\kappa$ is a dimensionless parameter which sets the scale of the function $k_1^2(\varphi)$.
Substituting into \cref{eq:constu}, one obtains
\begin{equation}
    \label{eq:uresult}
    \bar{u} = -\frac{2\kappa\Omega_\varphi}{n(1-\Omega_\varphi)}\,.
\end{equation}
The corresponding result in Ref.~\cite{wetterichinfquint} is given by
\begin{equation}
\label{eq:weturesult}
    \bar{u} = -\frac{2\kappa}{n(1-\Omega_\varphi)}\,,
\end{equation}
from which it follows that $\kappa$ must be small compared to $1$, in order to give a small $\bar{u}$ and hence produce behaviour close to the scaling solution. However, since $\Omega_\varphi$ is small, we find no such constraint on $\kappa$; $\bar{u}$ is small automatically in \cref{eq:uresult}.

This has implications for the prospects of constraining the model. A larger value of $\kappa$ gives smaller values of the function $k_1^2(\varphi)$ and hence smaller values of $\Omega_\varphi$~\cite{wetterichinfquint}. There is a tight upper bound from the Planck experiment on the value of $\Omega_\varphi$ at early times. This can translate into a lower bound on $\kappa$, which will be discussed in more detail in \cref{sec:results}. Based on \cref{eq:weturesult} one would conclude that there are both upper and lower bounds on $\kappa$, which could potentially put a very tight constraint on the model. However, based on our result for $\bar{u}$, which is small irrespective of the magnitude of $\kappa$, one finds no upper bound on $\kappa$. As will be shown in \cref{sec:results}, we can consider values of $\kappa$ much larger than the upper bound found in Ref.~\cite{wetterichinfquint}. Our numerical results match closely our prediction and there is no evidence of any approximation breaking down for large $\kappa$.

\subsection{Inverse power law potential}

An approximate analytic solution can also be found for models with inverse power law potentials of the form
\begin{equation}
    V(\varphi) = M_\text{P}^4 \Tilde{V}(M_\text{P}/\varphi)^\lambda\,,
\end{equation}
where $\tilde{V}$ and $\lambda$ are dimensionless constants.

While the neutrinos are relativistic, \cref{eq:SFE} becomes
\begin{equation}
\label{eq:SFE2}
\ddot{\varphi} + 3H\dot{\varphi} + \frac{1}{2k^2}\frac{\text{d}k^2}{\text{d}\varphi}\dot{\varphi}^2 + \frac{1}{k^2}\frac{\text{d}V}{\text{d}\varphi}  = 0\,.
\end{equation}
Using the same kinetic function as for the exponential potential case, \cref{eq:k}, but with $\alpha=1$ and $\bar{\varphi}=0$ since these parameters relate to the specific model presented in Ref.~\cite{wetterichinfquint}, we have
\begin{equation}
    k^2(\varphi) = \frac{M_\text{P}}{2\kappa\varphi}\,.
\end{equation}
For our choice of $k^2(\varphi)$ and $V(\varphi)$, \cref{eq:SFE2} becomes:
\begin{equation}
    \ddot{\varphi} + 3H\dot{\varphi} - \frac{\dot{\varphi}^2}{2\varphi} - 2\lambda\kappa M_\text{P}^{3+\lambda}\Tilde{V}\varphi^{-\lambda} = 0\,.
\end{equation}
In terms of e-foldings $N$ as the time variable, we have:
\begin{equation}
\label{eq:SFE2.5}
    H^2\varphi'' + HH'\varphi' + 3H^2\varphi' - \frac{H^2\varphi'^2}{2\varphi} - 2\lambda\kappa M_\text{P}^{3+\lambda}\Tilde{V}\varphi^{-\lambda} = 0\,,
\end{equation}
where $\varphi' \equiv \text{d}\varphi/\text{d}N$. Finally introducing $\Phi$ via 
\begin{equation}
\label{eq:Phiphi}
    \varphi = M_\text{P}\exp(\Phi/M_\text{P})\,,
\end{equation}
\cref{eq:SFE2.5} becomes:
\begin{multline}
\label{eq:SFE3}
    \Phi'' + \frac{\Phi'^2}{2M_\text{P}} + \left(\frac{H'}{H} + 3\right)\Phi' 
    \\- 2\lambda\kappa \frac{M_\text{P}^{3}\Tilde{V}}{H^2} \exp(-(\lambda+1)\Phi/M_\text{P}) = 0\,.
\end{multline}

In a radiation (matter) dominated universe, the Hubble parameter evolves according to $H^2 = \Tilde{H}^2\exp(-nN)$ where $n=4\,(3)$ and $\Tilde{H}$ is a normalising factor. \Cref{eq:SFE3} then becomes:
\begin{multline}
\label{eq:SFE4}
    \Phi'' + \frac{\Phi'^2}{2M_\text{P}} + \left(3-\frac{n}{2}\right)\Phi' 
    \\-  2\lambda\kappa\frac{M_\text{P}^{3}\Tilde{V}}{\Tilde{H}^2}\exp(nN-(\lambda+1)\Phi/M_\text{P}) = 0\,.
\end{multline}
Motivated by results from numerical simulation (see \cref{sec:results}), which show linear solutions for $\Phi$, we make the following ansatz:
\begin{equation}
\label{eq:ansatz}
    \Phi = qM_\text{P}N + \hat{\Phi}\,,
\end{equation}
where $q$ is a dimensionless constant and $\hat{\Phi}$
is the value $\Phi$ would take if this solution were extrapolated to $N=0$.

Under this ansatz \cref{eq:SFE4} becomes:
\begin{multline}
\label{eq:SFE5}
    \frac{1}{2}q^2M_\text{P} + \left(3-\frac{n}{2}\right)fM_\text{P} 
    \\- 2\lambda\kappa\frac{M_\text{P}^{3}\Tilde{V}}{\Tilde{H}^2}\exp(nN-(\lambda+1)qN - (\lambda+1)\hat{\Phi}/M_\text{P}) = 0\,.
\end{multline}
Treating the $N$-dependent and $N$-independent parts of the equation separately, we obtain 
\begin{equation}
\label{eq:grad}
    q = \frac{n}{\lambda+1} \,,
\end{equation}
which, substituting into \cref{eq:SFE5}, gives
\begin{multline}
    \frac{1}{2}\left(\frac{n}{\lambda+1}\right)^2 + \left(3-\frac{n}{2}\right)\frac{n}{\lambda+1} 
    \\- 2\lambda\kappa\frac{M_\text{P}^{2}\Tilde{V}}{\Tilde{H}^2}\exp(- (\lambda+1)\hat{\Phi}/M_\text{P}) = 0\,.
\end{multline}
Rearranging, we find $\hat{\Phi}$ as
\begin{multline}
\label{eq:Phi0}
    \hat{\Phi} = -\frac{M_\text{P}}{\lambda+1} \log\Biggl(\frac{\Tilde{H}^2}{2\lambda\kappa M_\text{P}^2\Tilde{V}}\Biggl( \frac{1}{2}\left(\frac{n}{\lambda+1}\right)^2 
    \\+ \left(3-\frac{n}{2}\right)\frac{n}{\lambda+1}\Biggr)\Biggr)\,.
\end{multline}

Thus, in contrast to the previous section, we find that inverse power law potentials admit solutions in which $\log(\varphi)$ evolves linearly with $N$ as opposed to $\varphi$ evolving linearly as in the exponential potential case. 

It is also instructive to find an expression for the dark energy density fraction.
Substituting our solution for $\varphi$ (\cref{eq:Phiphi,eq:ansatz}) into \cref{eq:rhophidef} gives the energy fraction:
\begin{multline}
    \Omega_\varphi = \frac{q^2}{12\kappa}\exp(qN+\hat{\Phi}/M_\text{P}) 
    \\+ \frac{M_\text{P}^{2}\tilde{V}}{3\Tilde{H}^2}\exp(nN-\lambda qN-\lambda\hat{\Phi}/M_\text{P})\,.
\end{multline}
Recalling \cref{eq:grad}, we can write
\begin{multline}
        \Omega_\varphi = \Biggl[\frac{q^2}{12\kappa}\exp(\hat{\Phi}/M_\text{P}) 
        \\+ \frac{M_\text{P}^{2}\tilde{V}}{3\Tilde{H}^2}\exp(-\lambda\hat{\Phi}/M_\text{P})\Biggr]\exp(qN)\,.
\end{multline}
Thus it turns out that $\Omega_\varphi$ is proportional to $\varphi$:
\begin{multline}
        \Omega_\varphi = \Biggl[\frac{q^2}{12\kappa}\exp(2\hat{\Phi}/M_\text{P}) 
        \\+ \frac{M_\text{P}^{2}\tilde{V}}{3\Tilde{H}^2}\exp((-\lambda+1)\hat{\Phi}/M_\text{P})\Biggr]\frac{\varphi}{M_\text{P}}\,,
\end{multline}
where $q$ and $\hat{\Phi}$ are given by \cref{eq:grad,eq:Phi0}. In contrast to the exponential case, where there is an approximately constant fraction of early dark energy, here the fact the dark energy fraction has an exponential dependence on $N$ implies that at early times (i.e. large negative values of $N$), it automatically makes a negligible contribution to the energy density.
These results are confirmed in \cref{sec:results}, with \cref{fig:logphis-N,fig:Omega_phi-N-log} showing $\log(\varphi)$ and $\log(\Omega_\varphi)$ evolving linearly with $N$ with a gradient given by $q$.

\section{Numerical evolution}
\label{sec:results}

In addition to the analytic approach laid out in \cref{sec:anal}, we numerically solved the equations of motion. This allows us to confirm the results of \cref{sec:anal} and to probe the late-universe cosmology that our analytic approach did not capture.

To generate our results we modified the code used by Barreira \textit{et al.} in Ref.~\cite{kmouflage}, which is based on the Code for Anisotropies in the Microwave Background (CAMB)~\cite{CAMB}. We modified the background part of Barreira \textit{et al.}'s code such that it solved the background equations of motion laid out in \cref{sec:model}.

We consider the following choices for the kinetic, potential and coupling functions:
\\
\\Kinetic function:
\begin{itemize}
    \item $k_\text{c}^2(\varphi) = \text{const}$\,,
    \item $k_1^2(\varphi) = \frac{M_\text{P}\alpha}{2\kappa(\varphi-\bar{\varphi})}$\,.
\end{itemize}
Coupling function:
\begin{itemize}
    \item $\beta_\text{c}(\varphi) = \text{const}$\,,
    \item $\beta_1(\varphi) = -  \frac{M_\text{P}}{\varphi_\text{c}-\varphi}$\,,
    \item $\beta_2(\varphi) = - \left( \frac{M_\text{P}}{\varphi_\text{c}-\varphi}\right)^2$\,,
    \item $\beta_3(\varphi) = - \frac{\gamma M_\text{P}}{\varphi}$\,.
\end{itemize}
Potential function:
\begin{itemize}
    \item $V_\text{exp}(\varphi) = M_\text{P}^4\exp(-\alpha\varphi/M_\text{P})$\,,
    \item $V_\text{IPL}(\varphi) = \tilde{V} M_\text{P}^4 (M_\text{P}/\varphi)^\lambda$\,.
\end{itemize}

The motivation for choosing these functions is as follows. In Ref.~\cite{wetterichinfquint}, the functions $k^2_1(\varphi)$ and $V_\text{exp}(\varphi)$ are used, with $\beta(\varphi)$ unspecified. We use this as a starting point, and we specify $\beta(\varphi)=\beta_1(\varphi)$ as employed in Ref.~\cite{Casas:2016duf}. We then widen the scope by choosing other functions that could be expected to give rise to growing neutrino quintessence behaviour. Inverse power law potentials
have a qualitatively similar `decaying' form to exponential potentials.
The couplings $\beta_\text{c}$, $\beta_1$, $\beta_2$, and $\beta_3$ each correspond to a function $C(\varphi)$ via \cref{eq:betadef}, or equivalently: 
\begin{equation}
    C(\varphi) = \exp\left(-\frac{1}{M_\text{P}}\int \beta(\tilde{\varphi})\text{d}\tilde{\varphi}\right)\,.
\end{equation}
The four functions $\beta(\varphi)$ considered here all correspond to a rapidly rising $C(\varphi)$. Thus $V(\varphi)$ and $C(\varphi)$ give rise to an effective potential for the scalar field that has a minimum, which is a necessary condition for growing neutrino quintessence.

In \cref{sec:exp} we focus on $k^2_1(\varphi)$, $\beta_1(\varphi)$ and $V_\text{exp}(\varphi)$. The scaling solution discussed in \cref{sec:anal} is verified and a constraint is found on the parameter $\kappa$ in $k_1^2(\varphi)$ due to its effect on the amount of early dark energy.
In \cref{sec:invpow} we consider $k^2_1(\varphi)$, $\beta_1(\varphi)$ and $V_\text{IPL}(\varphi)$, which give rise to qualitatively similar behaviour for the scalar field $\varphi$ but do not produce early dark energy.
We discuss the various options for $\beta(\varphi)$ in \cref{sec:beta}.

\subsection{Exponential potential}
\label{sec:exp}
\begin{figure}
    \centering
    \includegraphics[width=\columnwidth]{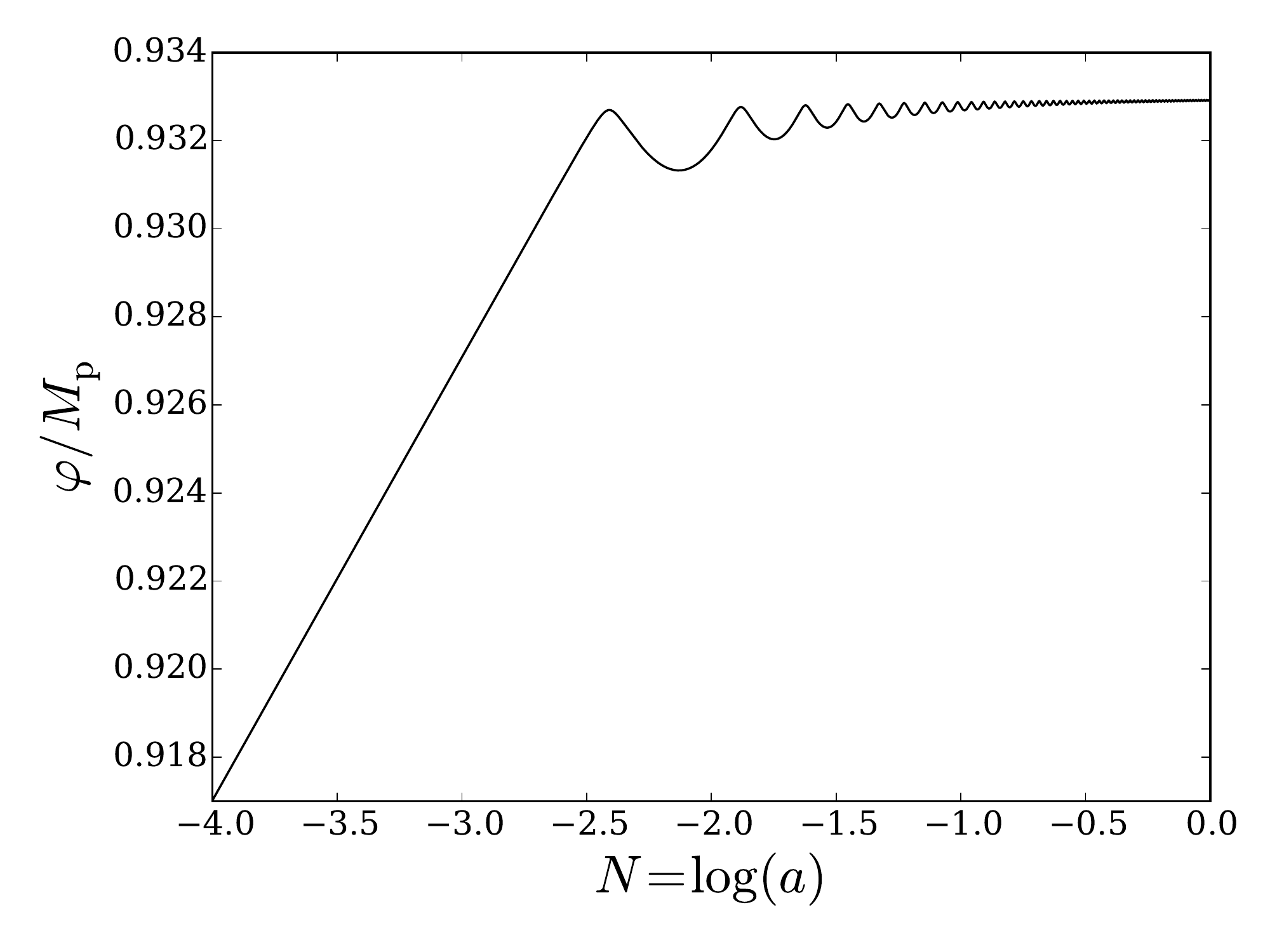}
    \caption{The late-time evolution of the scalar field for an exponential potential $V_\text{exp}(\varphi) = M_\text{P}^4\exp(-\alpha\varphi/M_\text{P})$, kinetic function $k^2_1(\varphi) = M_\text{P}\alpha/(2\kappa(\varphi-\bar{\varphi}))$ and coupling function $\beta_1(\varphi) = -  M_\text{P}/(\varphi_\text{c}-\varphi)$  with $\alpha = 300$, $\kappa=1.8$ $\bar{\varphi}=0.0933$ and $\varphi_\text{c}=0.933$.}
    \label{fig:phizoom}
\end{figure}

In this section we present the results of numerical calculations using $V_\text{exp}(\varphi)$, $k^2_1(\varphi)$ and $\beta_1(\varphi)$. During radiation and matter domination we find $\varphi$ evolving linearly with $N=\log(a)$ according to the scaling solution discussed in \cref{sec:anal}. After the neutrinos become non-relativistic, $\varphi$ starts to oscillate around the minimum of the effective potential formed by $V(\varphi)$ and $\beta(\varphi)$ and comes to a halt to behave as an effective cosmological constant. This behaviour is illustrated in \cref{fig:phizoom}.

\begin{figure}
    \centering
    \includegraphics[width=\columnwidth]{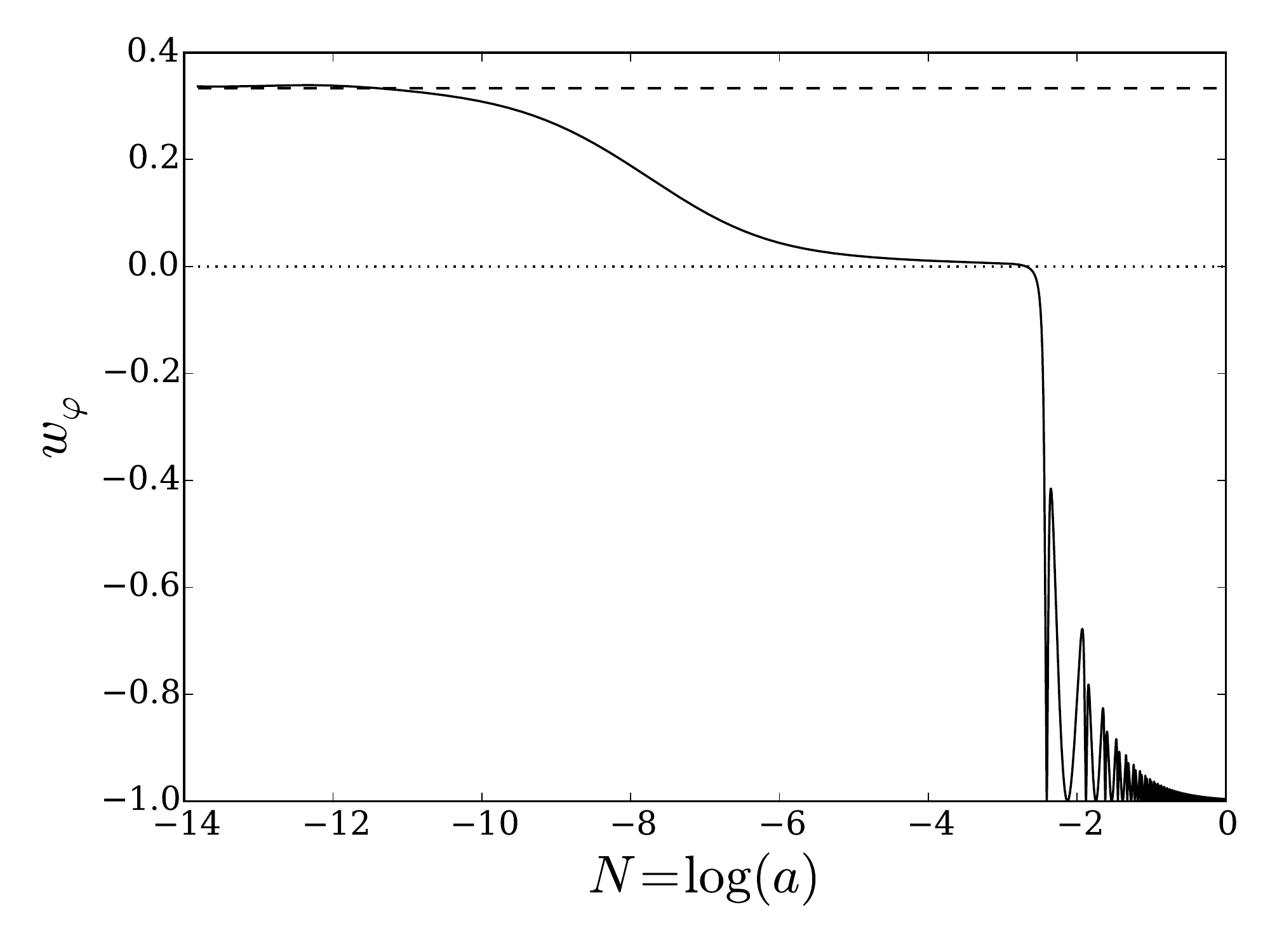}
    \caption{The evolution of the equation of state of the scalar field, $w_\varphi$, for the same functions and parameters as in \cref{fig:phizoom}. The dashed and dotted lines show the equation of state during radiation and matter domination respectively.}
    \label{fig:w_phi-N}
\end{figure}

\begin{figure}
    \centering
    \includegraphics[width=\columnwidth]{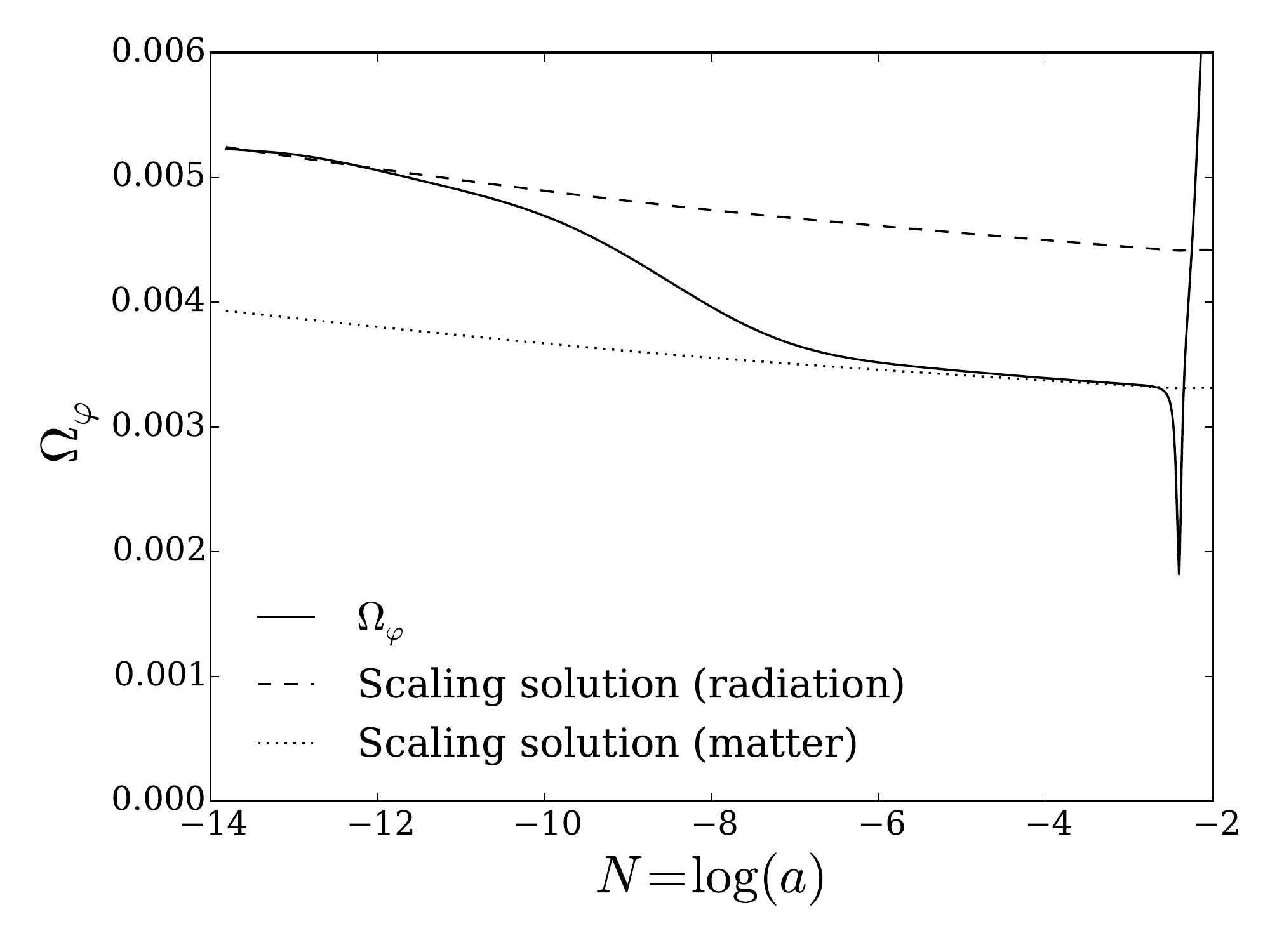}
    \caption{The evolution of the energy density fraction of the scalar field, $\Omega_\varphi$ during radiation and matter domination (solid line) for the same functions and parameters as in \cref{fig:phizoom}. The dashed and dotted lines respectively show the predicted evolution of $\Omega_\varphi$ under the assumption of a radiation dominated and matter dominated universe where the scalar field obeys the scaling solution discussed in \cref{sec:anal}.}
    \label{fig:Omega_phi-N-3}
\end{figure}

\Cref{fig:w_phi-N} shows the evolution of the equation of state of the scalar field. It can be seen that it mimics radiation with a value of $w_\varphi=1/3$ when the Universe is radiation dominated, then approaches $w_\varphi=0$, mimicking matter when the Universe is matter dominated, and finally tends towards $w_\varphi=-1$ after the neutrinos halt the evolution of the scalar field and it mimics a cosmological constant. The first two regimes illustrate the scaling solution, where the energy density of the scalar field tracks that of the dominant species as discussed in \cref{sec:anal}. This is also illustrated in \cref{fig:Omega_phi-N-3}, in which we have plotted the predictions of the energy density fraction of the scalar field assuming the scaling solution is exactly satisfied both for radiation and matter domination. It can be seen that in the early Universe the numerical result closely follows $\Omega_\varphi = 4k^2(\varphi)/\alpha^2$ and at later times it follows  $\Omega_\varphi = 3k^2(\varphi)/\alpha^2$, with a transition in between, as expected.

\begin{figure}
    \centering
    \includegraphics[width=\columnwidth]{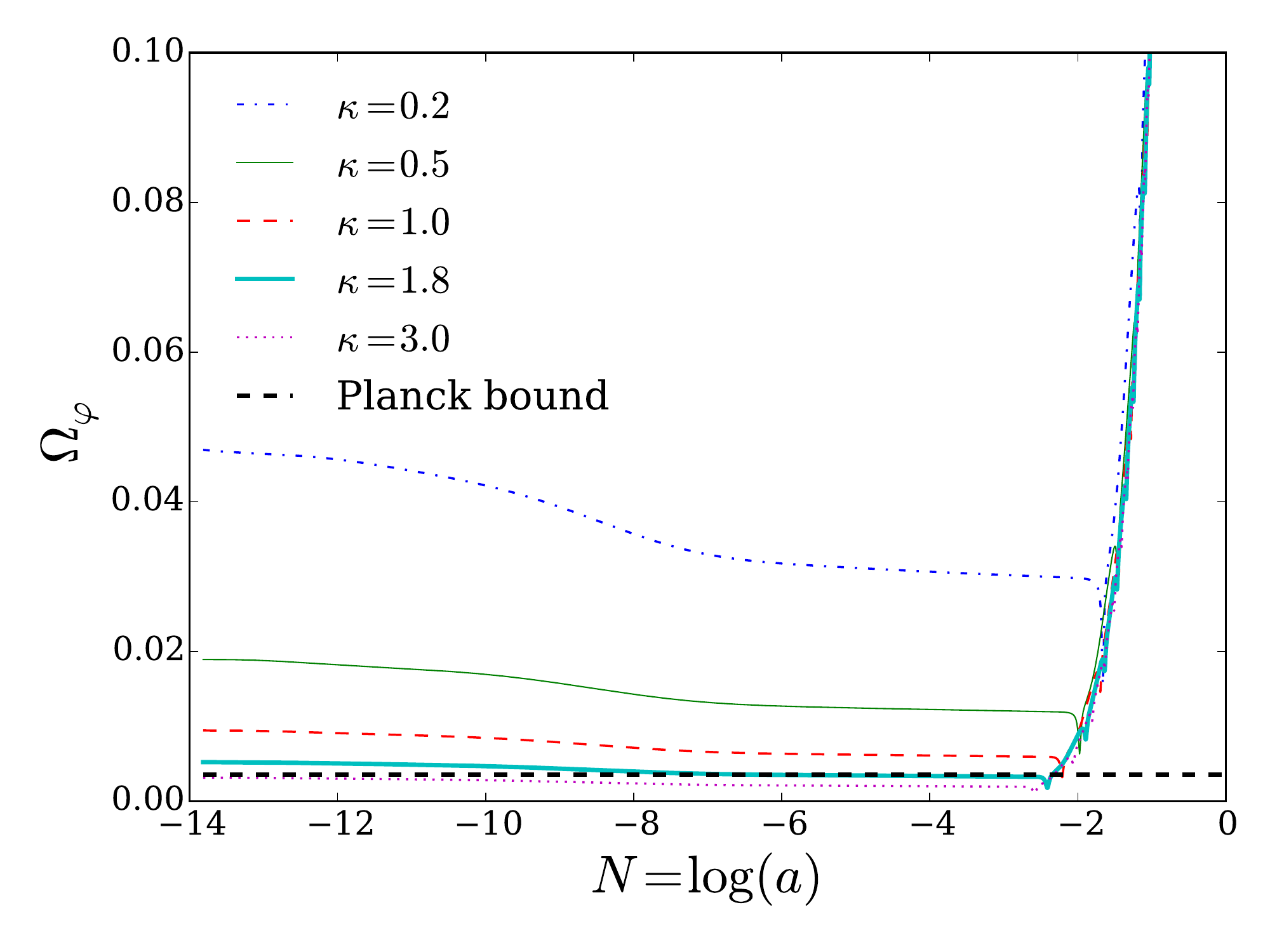}
    \caption{The evolution of the energy density fraction of the scalar field for a range of values of $\kappa$ and otherwise the same functions and parameters as in \cref{fig:phizoom}. Also shown is the Planck upper bound on $\Omega_\text{e}<0.0036$.}
    \label{fig:kappazoom}
\end{figure}

\Cref{fig:kappazoom} shows the effect of varying the model parameter $\kappa$ in $k_1^2(\varphi)$ on the scalar field evolution and the energy density fraction of the scalar field respectively. Note that the larger the value of $\kappa$ the smaller the amount of early dark energy. This agrees with the scaling solution result, \cref{eq:Oscalsol} in \cref{sec:anal}, since $\kappa$ is effectively a constant that controls the size of the kinetic function $k^2_1(\varphi)$ as can be seen in \cref{eq:k}. 

We find that our numerical results for the evolution of dark energy are well approximated by the early dark energy parametrisation of Doran \& Robbers~\cite{DoranRobbers}, in which the dark energy density fraction is parametrised as follows: 
\begin{equation}
    \Omega_\text{DE}(a) = \frac{\Omega_\text{DE}^0-\Omega_\text{e}(1-a^{-3w_0})}{\Omega_\text{DE}^0 + \Omega_\text{m}^0a^{3w_0}} + \Omega_\text{e}(1-a^{-3w_0})\,,
\end{equation}
where $\Omega_\text{e}$ (the fraction of early dark energy) and $w_0$ (the present day equation of state) are parameters to be fitted and $\Omega_\text{DE}^0$ and $\Omega_\text{m}^0$ are the present-day dark energy and matter fractions. For a given value of $\kappa$ we carry out a least-squares fitting of our numerical results to the Doran \& Robbers parametrisation to give $w_0$ and $\Omega_\text{e}$. 
The Planck Collaboration~\cite{PlanckXIV} finds an upper bound on the parameter $\Omega_\text{e}$ of $0.0036$. We find that the value of $\kappa$ required to give rise to this value of $\Omega_\text{e}$ is $\kappa=1.8$, with larger values of $\kappa$ resulting in smaller values of $\Omega_\text{e}$ and vice versa. We therefore find a lower bound on $\kappa$ of $1.8$. 

As discussed in \cref{sec:anal}, Ref.~\cite{wetterichinfquint} finds a requirement that $\kappa \ll 1$ in order to ensure that $u$, the deviation of $\Omega_\varphi$ from the scaling solution at early times, is small. If this requirement were valid then the model of Ref.~\cite{wetterichinfquint} would have been ruled out by the constraints on early dark energy. However, due to our finding in \cref{sec:anal} that $u$ is given by \cref{eq:uresult} and not \cref{eq:weturesult}, we find that there is no requirement for $\kappa$ to be small and hence our constraint that $\kappa>1.8$ does not rule out the model.

Varying the parameter $\alpha$ in the potential merely results in a rescaling of $\varphi$ and does not have any effect on the physics.
We also studied the case of a constant kinetic function $k^2_\text{c}$ and found that it made almost no difference to the results. This is as expected, because the varying kinetic function to which we compare it varies very slowly over the relevant part of the Universe's history. 

\subsection{Inverse power law potential}
\label{sec:invpow}
\begin{figure}
    \centering
    \includegraphics[width=\columnwidth]{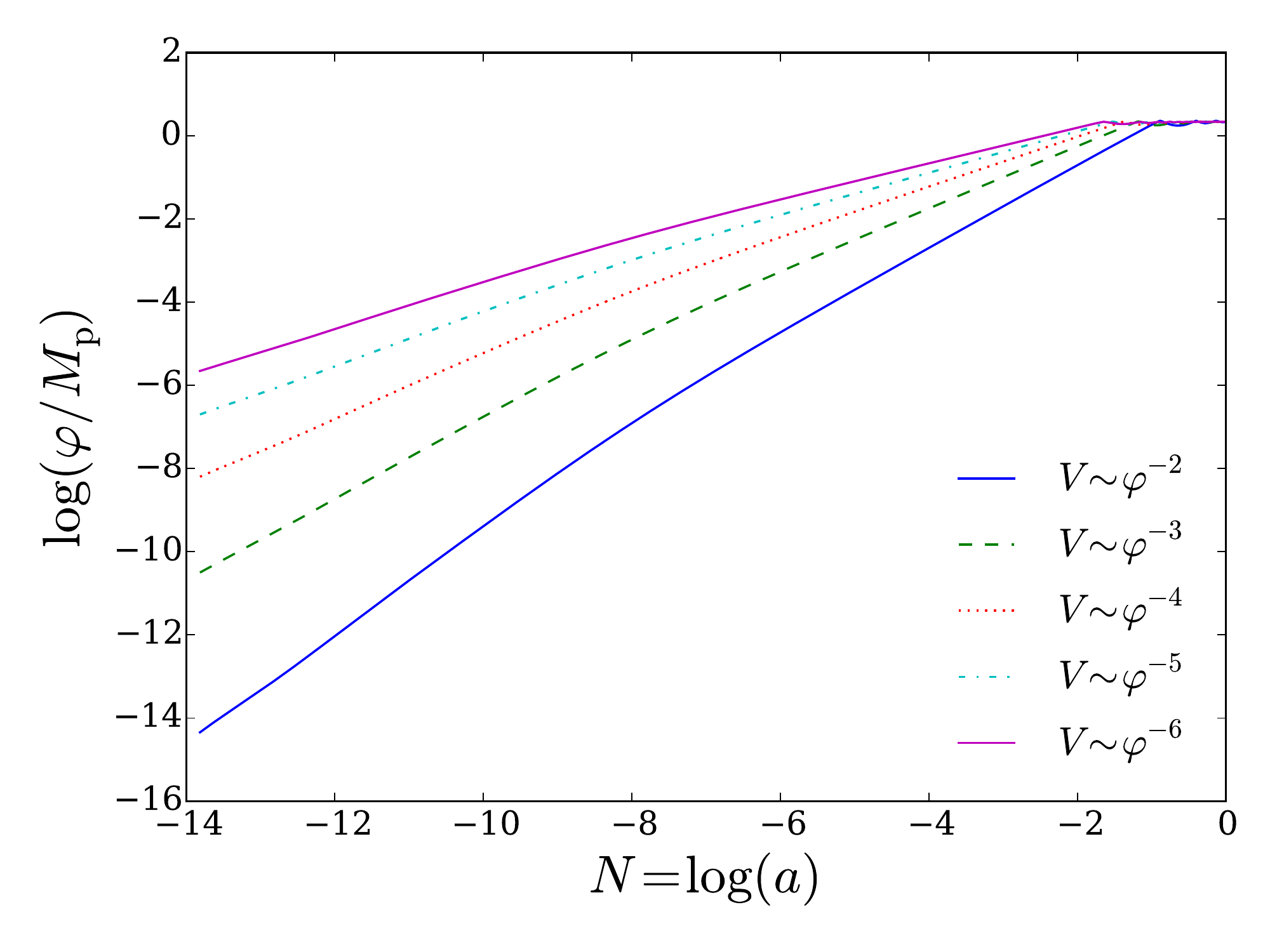}
    \caption{The evolution of the logarithm of the scalar field for inverse power law potentials of the form $V_\text{IPL}(\varphi) = \tilde{V} M_\text{P}^4 (M_\text{P}/\varphi)^\lambda$ with kinetic function $k_1^2(\varphi)=M_\text{P}/(2\kappa\varphi)$ and coupling function $\beta_1(\varphi)=-M_\text{P}/(\varphi_\text{c}-\varphi)$. We fix $\kappa=1.8$ and the parameters $\tilde{V}$ and $\varphi_\text{c}$ take different values for different values of $\lambda$ (see text for details).}
   \label{fig:logphis-N}
\end{figure}

\begin{figure}
    \centering
    \includegraphics[width=\columnwidth]{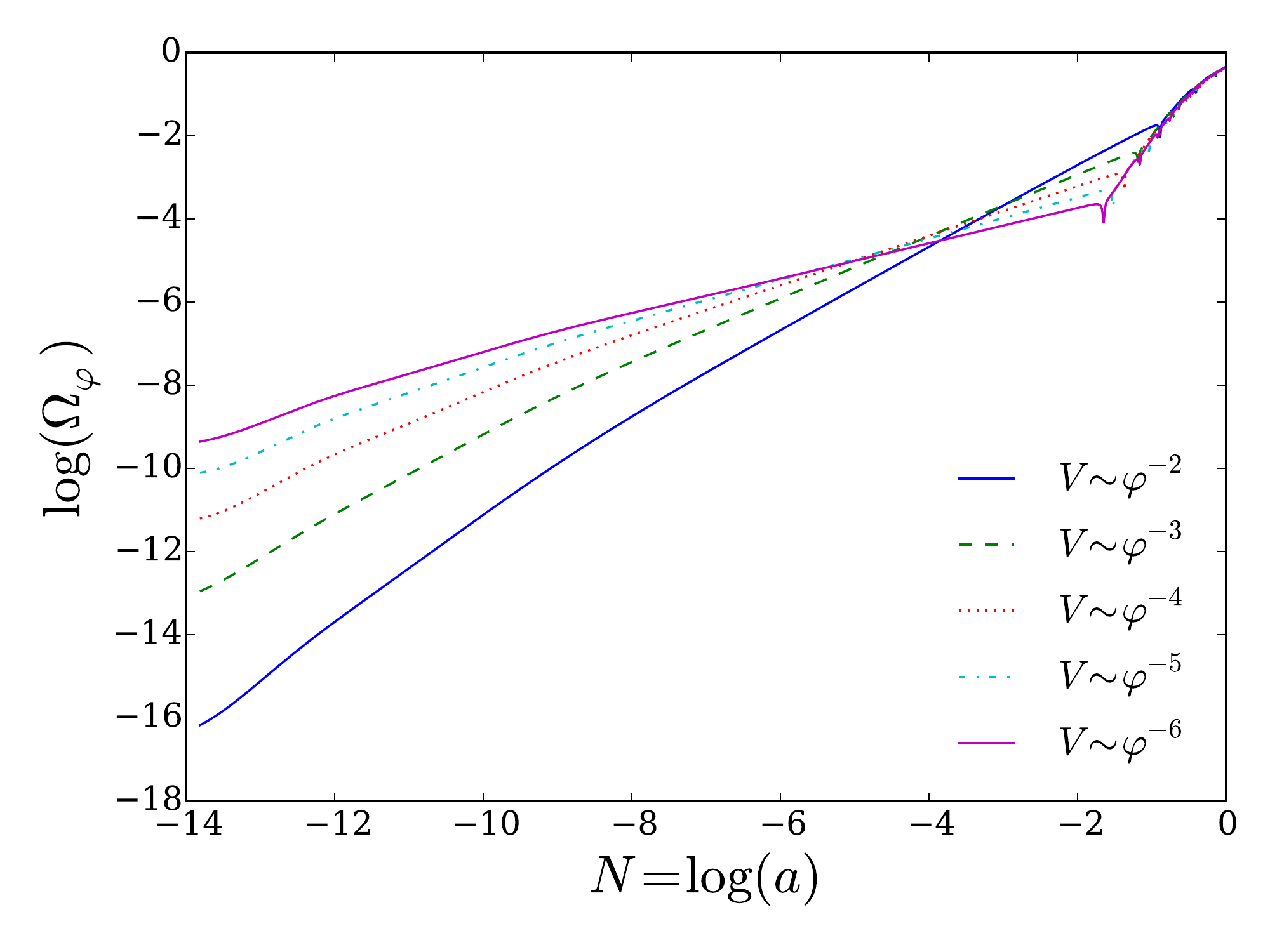}
    \caption{The evolution of the logarithm of the energy fraction of the scalar field for the same functions and parameters as in \cref{fig:logphis-N}.}
    \label{fig:Omega_phi-N-log}
\end{figure}

In this section we present the results for models with $V_\text{IPL}(\varphi)$, $k^2_1(\varphi)$ and $\beta_1(\varphi)$, with $\kappa=1.8$, $\alpha=1$ and $\bar{\varphi}=0$. We considered several different values of the power $\lambda$ as shown in \cref{fig:logphis-N,fig:Omega_phi-N-log}. For each value of $\lambda$, an appropriate value of $\tilde{V}$ was chosen to produce the correct dark energy density fraction at the present day. For ease of comparison, the same present-day value of $\varphi$ was chosen for each value of $\lambda$, with $\varphi_\text{c}$ being tuned in each case to achieve this.

The choice of $\kappa=1.8$ was made for ease of comparison with the exponential potential, but has no special significance in the inverse power law case. Larger values of $\kappa$ result in an upward shift in $\varphi$ and a corresponding downward shift in $\Omega_\varphi$.

Compared to the models with exponential potentials already discussed, the behaviour of models with inverse power law potentials is not drastically different. During radiation and matter domination we find that $\varphi$ evolves exponentially with $N$ as opposed to linearly as it does for models with $V_\text{exp}(\varphi)$. However,
the qualitative behaviour of the field increasing as long as neutrinos are relativistic and then effectively stopping once they become non-relativistic is still present. 
\Cref{fig:logphis-N} shows the evolution of the logarithm of the scalar field against $N$ for different inverse power law potentials. Before the neutrinos become non-relativistic, $\log(\varphi)$ evolves approximately linearly with a gradient of $n/(\lambda+1)$ and an intercept of $\hat{\Phi}$ as predicted in \cref{eq:grad,eq:Phi0}.

The evolution of the energy density of the scalar field is shown in \cref{fig:Omega_phi-N-log}. From this it is clear that these models do not give rise to early dark energy; looking back in time, the energy density of the scalar field continues to drop off rapidly. 
The constraint on $\kappa$ that we found for exponential potentials therefore does not apply to models with inverse power law potentials.

\subsection{Coupling function}
\label{sec:beta}

In addition to the coupling $\beta_1(\varphi)$ already considered, we investigated $\beta_\text{c}$, $\beta_2(\varphi)$ and $\beta_3(\varphi)$. None of these choices led to behaviour significantly different to the $\beta_1(\varphi)$ case, provided $|\beta|$ is sufficiently large at the time at which neutrinos become non-relativistic. This requirement is automatically satisfied for $\beta_1(\varphi)$ and $\beta_2(\varphi)$, since as $\varphi$ approaches $\varphi_\text{c}$, $|\beta(\varphi)|$ tends to infinity. The scalar field is never allowed to reach $\varphi_\text{c}$, however, because the neutrino coupling term in the scalar field equation, \cref{eq:SFE}, always acts to decrease the value of $\varphi$. It can be seen that the value of $\varphi_\text{c}$ in $\beta_1(\varphi)$ and $\beta_2(\varphi)$ determines the present-day value of $\varphi$, since the latter will approach ever closer to it but can never exceed it. This is confirmed by our numerical analysis.

For $\beta_\text{c}$ and $\beta_3(\varphi)$ one does not automatically obtain large $|\beta|$ but it must be set by an appropriate choice of parameters. In the latter case this means choosing a large value of $\gamma$. 
The requirement on the size of $|\beta|$ is illustrated by the following result from Ref.~\cite{wetterichvargrav}:
\begin{equation}
\label{eq:betamag}
    \frac{\Omega_\varphi}{\Omega_\nu} \approx |\beta|\,.
\end{equation}
If $|\beta|$ is too small, the coupling term in \cref{eq:SFE} will not be large enough to counteract the potential term and the value of $\varphi$ will continue to increase. This will result in both a larger $\Omega_\nu$ and a smaller $\Omega_\varphi$. Our numerical results are consistent with \cref{eq:betamag}.

\section{Conclusions}
\label{sec:conc}

We have considered various growing neutrino quintessence scenarios inspired by the model proposed by Wetterich in Ref.~\cite{wetterichinfquint}. We studied the early-universe background evolution analytically for both exponential potentials and inverse power law potentials. In the former case we followed the procedure in Ref.~\cite{wetterichinfquint}, finding some disagreement with their results. In the latter case we found an analytic result for the behaviour of such models that we were able to check numerically.

Using a modified version of CAMB, we found the numerical solution to the background equations for several different kinetic, potential and neutrino-scalar coupling functions. We verified our analytic predictions, investigated the circumstances under which growing neutrino quintessence behaviour is obtained, and used early dark energy constraints~\cite{PlanckXIV} to constrain the parameter $\kappa$ that controls the scale of the kinetic function $k^2_1(\varphi)$.

The following conditions must be met to give rise to growing neutrino quintessence:
\begin{itemize}
    \item $V(\varphi)$ must have a negative gradient in order to cause the value of the scalar field to increase with time. This gradient must be sufficiently steep that $\varphi$ reaches large enough values in the late Universe to act as dark energy. Note that growing neutrino quintessence models such as the ones considered here do not require that $V(\varphi)$ be flat in the late Universe, as other quintessence models often require. The slow evolution of $\varphi$ necessary for it to mimic a cosmological constant is achieved by the presence of the neutrino coupling term, not by slow roll.
    \item $|\beta(\varphi)|$ must be sufficiently large when the neutrinos become non-relativistic that $\beta(\rho_\nu-3p_\nu)$ is able to act as a strong enough restoring force to stop the evolution of $\varphi$ in \cref{eq:SFE}.
\end{itemize}

For exponential potentials with slowly varying kinetic functions we found the predicted scaling solution behaviour with an approximately constant early dark energy fraction. Using existing constraints on early dark energy~\cite{PlanckXIV} we were able to constrain the model parameter $\kappa$, which sets the scale of the kinetic function, to be larger than $1.8$, forcing it into a region previously thought excluded~\cite{wetterichinfquint}. However, we also found 
that there is no upper bound on $\kappa$ and so the model is not ruled out.

As well as the exponential potential considered in Ref.~\cite{wetterichinfquint}, we have also considered inverse power law potentials, since these have a qualitatively similar form to exponential potentials and so could provide the necessary conditions for growing neutrino quintessence. 
We confirmed that such models can give rise to growing neutrino quintessence and we found that, unlike in the case of exponential potentials, there is no early dark energy present. Planck bounds on early dark energy therefore do not translate into constraints on models with inverse power law potentials.

\section{Acknowledgements}
FNC is supported by a United Kingdom Science and Technology Facilities Council (STFC) studentship. AA, EJC and AMG acknowledge  support  from  STFC  grant ST/P000703/1. BL is supported by the European Research Council via grant ERC-StG-716532-PUNCA, and by STFC Consolidated Grants ST/P000541/1, ST/L00075X/1. We are grateful to Marco Baldi and Christoph Wetterich for useful discussions  and to the referee for helpful and detailed comments on an earlier version of this manuscript.

\appendix
\section{}
\label{appendix}
In this Appendix we present our derivation of the results presented in \cref{sec:anal}. The derivation follows that in Ref.~\cite{wetterichinfquint}, but we find a few disagreements with their results.

In the scaling solution, the energy density of the scalar field is a constant fraction of that of the dominant species:
\begin{equation}
    \rho_\varphi = f \rho_\text{d}\,.
\end{equation}
To find solutions close to the scaling solution we allow $f$ to vary as a function of $\varphi$:
\begin{equation}
\label{eq:fdef}
\rho_\varphi = f(\varphi)\rho_\text{d}\,,
\end{equation}
and allow a small deviation $\delta(N)$ from the scaling solution result for $\varphi$ (\cref{eq:phiscalsol}):
\begin{equation}
\label{eq:deltadef}
    \varphi = M_\text{P}\frac{nN}{\alpha} + \hat{\varphi} + M_\text{P}\delta(N)\,.
\end{equation}

Differentiating \cref{eq:fdef}, we find:
\begin{equation}
\label{eq:logfprime}
    (\log f)' =  (\log \rho_\varphi)' -  (\log \rho_\text{d})'\,,
\end{equation}
where primes denote differentiation with respect to $N$.
It will be necessary to employ the $\rho_\varphi$ conservation equation, \cref{eq:rhophicons} (with $p_\nu=\rho_\nu/3$), as well as the definitions of $\rho_\varphi$ and $p_\varphi$, \cref{eq:rhophidef,eq:pphidef}. Using $N$ as the time variable, these are given by:
\begin{equation}
\label{eq:rhophicons2}
    \rho_\varphi' = -3(\rho_\varphi + p_\varphi)\,,
\end{equation}
\begin{equation}
\label{eq:rhophidef2}
    \rho_\varphi = \frac{k^2H^2}{2}\varphi'^2 + V\,,
\end{equation}
and
\begin{equation}
\label{eq:pphidef2}
    p_\varphi = \frac{k^2H^2}{2}\varphi'^2 - V\,.
\end{equation}
It proves convenient to introduce the constant of proportionality in \cref{eq:rhod} as follows:
\begin{equation}
\label{eq:rhobardef}
    \rho_\text{d} = \bar{\rho} M_\text{P}^4 \exp(-nN-\alpha\hat{\varphi}/M_\text{P})\,.
\end{equation}
Substituting \cref{eq:rhophicons2,eq:rhophidef2,eq:pphidef2,eq:rhobardef} into \cref{eq:logfprime} yields
\begin{equation}
\label{eq:dlnf1}
    (\log f)' = -6\left(1-\frac{V}{\rho_\varphi}\right) + n\,.
\end{equation}
Now, using \cref{eq:fdef,eq:deltadef,eq:rhobardef}
\begin{equation}
    \label{eq:V/rho}
    \frac{V}{\rho_\varphi} = \frac{M_\text{P}^4 \exp(-\alpha\varphi/M_\text{P})}{f \bar{\rho} M_\text{P}^4 \exp(-nN-\alpha\hat{\varphi}/M_\text{P})} = \frac{1}{f\bar{\rho}}\exp(-\alpha\delta)\,.
\end{equation}
Hence \cref{eq:dlnf1} becomes:
\begin{equation}
\label{eq:dlnf2}
    (\log f)' = n -6 + \frac{6}{f \bar{\rho}} \exp(-\alpha\delta)\,.
\end{equation}
Differentiating \cref{eq:deltadef} gives
\begin{equation}
\label{eq:ddelta1}
    \delta' = \frac{\varphi'}{M_\text{P}} - \frac{n}{\alpha}\,.
\end{equation}
Rearranging \cref{eq:rhophidef2}, we write $\varphi'$ in terms of $\rho_\varphi$ and $V$:
\begin{equation}
    \varphi' = \left[\frac{2\rho_\varphi}{k^2H^2}\left(1 - \frac{V}{\rho_\varphi}\right)\right]^\frac{1}{2}\,,
\end{equation}
and hence
\begin{equation}
    \varphi' = M_\text{P}\left[\frac{6\Omega_\varphi}{k^2}\left(1 - \frac{1}{f\bar{\rho}}\exp(-\alpha\delta)\right)\right]^\frac{1}{2}\,,
\end{equation}
where we have used \cref{eq:V/rho} again. Substituting into \cref{eq:ddelta1} yields
\begin{equation}
\label{eq:ddelta2}
    \delta' = -\frac{n}{\alpha} + \left[\frac{6\Omega_\varphi}{k^2}\left(1 - \frac{1}{f\bar{\rho}}\exp(-\alpha\delta)\right)\right]^\frac{1}{2}\,.
\end{equation}
For constant $k^2$ the scaling solution is recovered, with $f=\text{const}$ and $\delta=0$. \Cref{eq:dlnf2} then gives
\begin{equation}
\label{eq:fscal}
    \frac{1}{f} = \left(1-\frac{n}{6}\right)\bar{\rho}\,.
\end{equation}
If, however, $k^2$ varies smoothly we expect only a small deviation from this solution. We introduce a function $\zeta(N)$ to quantify the deviation of $f$ from the scaling solution value given by \cref{eq:fscal}:
\begin{equation}
\label{eq:zetadef}
    \frac{1}{f} = \left(1-\frac{n}{6}\right)\bar{\rho}\exp(-\alpha\zeta)\,.
\end{equation}

Differentiating \cref{eq:zetadef} gives
\begin{equation}
    \zeta' = \frac{1}{\alpha} (\log f)'\,,
\end{equation}
which, using \cref{eq:dlnf2}, gives
\begin{equation}
\label{eq:dzeta0}
    \zeta' = \frac{1}{\alpha}\left[n -6 + \frac{6}{f \bar{\rho}} \exp(-\alpha\delta)\right]\,.
\end{equation}
\Cref{eq:dzeta0,eq:ddelta2}, both contain the term $1/(f\bar{\rho})\exp(-\alpha\delta)$, which using \cref{eq:zetadef} can be written as
\begin{equation}
    \label{eq:frhoexpdelta}
    \frac{1}{f\bar{\rho}}\exp(-\alpha\delta) = \left(1-\frac{n}{6}\right)\exp(-\alpha(\delta+\zeta))\,.
\end{equation}
\Cref{eq:dzeta0,eq:ddelta2} can now be written as
\begin{multline}
\label{eq:ddelta3}
    \delta' = -\frac{n}{\alpha} + 
    \\\left[\frac{n\Omega_\varphi}{k^2}\right]^{\frac{1}{2}} \left[1+\left(\frac{6}{n}-1\right)\left(1-\exp[-\alpha(\delta+\zeta)]\right)\right]^\frac{1}{2}\,,
\end{multline}
and
\begin{equation}
\label{eq:dzeta1}
    \zeta' = \frac{n-6}{\alpha}\left[1-\exp(-\alpha(\delta+\zeta))\right]\,
\end{equation}
respectively.

Now we recall \cref{eq:Oscalsol}, but introduce a small deviation $u(N)$, by
\begin{equation}
\label{eq:udef}
    \Omega_\varphi  = \frac{nk^2}{\alpha^2}(1-u)\,,
\end{equation}
and group the small functions $\delta$ and $\zeta$ through the small function $\Delta$, defined by
\begin{equation}
\label{eq:Deltadef}
    \Delta = \left(\frac{6}{n}-1\right)\left(1-\exp[-\alpha(\delta+\zeta)]\right)\,.
\end{equation}
Differentiating \cref{eq:Deltadef}, we find
\begin{equation}
    \label{eq:dDelta1}
    \Delta' = \alpha\left(\frac{6}{n}-1\right)\exp[-\alpha(\delta+\zeta)](\delta' + \zeta')\,.
\end{equation}
We can make use of \cref{eq:udef,eq:Deltadef} to simplify our equations for $\delta'$ and $\zeta'$, \cref{eq:ddelta3,eq:dzeta1} as follows:
\begin{equation}
    \label{eq:ddelta4}
    \delta' = \frac{n}{\alpha}\left(\sqrt{(1-u)(1+\Delta)}-1\right)\,,
\end{equation}
\begin{equation}
    \label{eq:dzeta2}
    \zeta' = -\frac{n}{\alpha}\Delta\,.
\end{equation}
Substituting \cref{eq:ddelta4,eq:dzeta2,eq:Deltadef} into \cref{eq:dDelta1} gives
\begin{equation}
\label{eq:dDelta2}
    \Delta' = [6-n(1+\Delta)](\sqrt{(1-u)(1+\Delta)} - 1 - \Delta)\,.
\end{equation}

The differential equation for $u$ follows from differentiating \cref{eq:udef} and rearranging as
\begin{equation}
\label{eq:du1}
    u' = (1-u)\left[\frac{\text{d}\log k^2}{\text{d}\varphi}\varphi' - (\log\Omega_\varphi)'\right]\,,
\end{equation}
which in turn yields
\begin{equation}
    \label{eq:du2}
    u' = (1-u)\left[M_\text{P}\frac{\text{d}\log k^2}{\text{d}\varphi}\left(\frac{n}{\alpha} + \delta'\right) - \frac{\alpha}{1+f}\zeta'\right]\,,
\end{equation}
where we have made use of \cref{eq:ddelta1,eq:dzeta1} and the fact that $\Omega_\varphi = f/(1+f)$. \Cref{eq:du2,eq:dDelta2} can be compared to Eq.~(108) in Ref.~\cite{wetterichinfquint}. We find two instances of the factor $(1-u)$ instead of $(1+u)$, and the second term in \cref{eq:du2} differs by a factor of $\Omega_\varphi$. This latter difference follows through to give an extra factor of $\Omega_\varphi$ in \cref{eq:uresult} compared to Ref.~\cite{wetterichinfquint} which, as discussed in \cref{sec:anal}, has a crucial impact on the range of possible values for the parameter $\kappa$.

We continue following the procedure of Ref.~\cite{wetterichinfquint} but with our versions of the $\Delta$ and $u$ equations in order to find an approximate form for $u$. Using \cref{eq:ddelta4,eq:dzeta2}, \cref{eq:du2} can be rewritten as
\begin{equation}
\label{eq:du3}
    u' = (1-u)\left[\frac{M_\text{P}n}{\alpha}\frac{\text{d}\log k^2}{\text{d}\varphi}\sqrt{(1-u)(1+\Delta)} + \frac{n}{1+f}\Delta\right]\,.
\end{equation}
Close to the scaling solution $\Delta$, $u$ and $M_\text{P}\text{d}\log k^2/\text{d}\varphi$ are all small. 
Expanding \cref{eq:dDelta2,eq:du3} to linear order in small quantities gives
\begin{align}
\Delta' &= \frac{n-6}{2}(\Delta + u)\notag
\\u' &= \frac{nM_\text{P}}{\alpha}\frac{\text{d}\log k^2}{\text{d}\varphi} + n(1-\Omega_\varphi)\Delta\,.
\end{align}
Setting $\Delta'= u'=0$, we see that this system of equations admits a constant solution:
\begin{equation}
\label{eq:ubar}
    \bar{u} = -\bar{\Delta} = \frac{M_\text{P}}{\alpha(1-\Omega_\varphi)}\frac{\text{d}\log k^2}{\text{d}\varphi}\,.
\end{equation}
One can then split $u = \bar{u} + \hat{u}$ and $\Delta = \bar{\Delta} + \hat{\Delta}$ into their $N$-independent and $N$-dependent components. The equations of motion for only the $N$-dependent parts are as follows:
\begin{align}
    \label{eq:uDelta}
    \hat{\Delta}' &= \frac{n-6}{2}(\hat{\Delta}+\hat{u})
    \\ \hat{u}' &= n(1-\Omega_\varphi)\hat{\Delta}\,,
\end{align}
which can be written in the following form:
\begin{equation}
\begin{pmatrix}
\hat{\Delta}\\\hat{u}
\end{pmatrix}'
= A
\begin{pmatrix}
\hat{\Delta}\\\hat{u}
\end{pmatrix}\,,
\end{equation}
where
\begin{equation}
    A = \frac{n-6}{2}
    \begin{pmatrix}
    1 & 1 \\\frac{2n(1-\Omega_\varphi)}{n-6} & 0
    \end{pmatrix}\,.
\end{equation}
The real parts of the eigenvalues of the matrix $A$ are both negative, which implies that the $N$-dependent parts of $\Delta$ and $u$ decay with $N$. Thus the solution with $u=\bar{u}$ and $\Delta = \bar{\Delta}$ is approached. Hence, it is appropriate to use $\bar{u}$ in \cref{eq:udef} which gives rise to \cref{eq:EDE1,eq:constu}.

\bibliography{general.bib}

\end{document}